# May oxygen depletion explain the FLASH effect? A chemical track structure analysis

Daria Boscolo[a], Emanuele Scifoni[b], Marco Durante[a,c]*, Michael Krämer[a], and Martina C. Fuss[a]*

[a]GSI Helmholtzzentrum für Schwerionenforschung GmbH, 64291 Darmstadt, Germany.
[b]Trento Institute for Fundamental Physics and Applications (TIFPA), National Institute for Nuclear Physics (INFN), 38123 Trento, Italy. [c]Institut für Physik Kondensierter Materie, Technische Universität Darmstadt, 64289 Darmstadt, Germany

* Corresponding authors: Marco Durante, Martina C. Fuss
Planckstr. 1
64291 Darmstadt
Germany
Email:  m.durante@gsi.de, m.fuss@gsi.de

**Keywords**

FLASH effect mechanism, radiation track structure simulation, oxygen depletion, reactive oxygen species

**Highlights**

- Bottom-up radiation chemistry study reproduced observed radiolytic oxygen depletion (ROD) with high accuracy
- Dynamical nature of oxygen depletion and its impact on OER taken into account for the first time
- Negligible impact of ROD on radiosensitivity through transient hypoxia in conditions of reported experiments
- ROD impact on therapeutic window occurs eventually in opposite direction

**Abstract**

**Background and purpose:** Recent observations in animal models show that ultra-high dose rate ("FLASH") radiation treatment significantly reduces normal tissue toxicity maintaining an equivalent tumor control. The dependence of this "FLASH" effect on target oxygenation has led to the assumption that oxygen "depletion" could be its major driving force.

**Materials and Methods:** In a bottom-up approach starting from the chemical track evolution of 1 MeV electrons in oxygenated water simulated with the TRAX-CHEM Monte Carlo code, we determine the oxygen consumption and radiolytic reactive oxygen species production following a short radiation pulse. Based on these values, the effective dose weighted by oxygen enhancement ratio (OER) or the *in vitro* cell survival under dynamic oxygen pressure is calculated and compared to that of conventional exposures, at constant OER.

**Results:** We find an excellent agreement of our Monte Carlo predictions with the experimental value for radiolytic oxygen removal from oxygenated water. However, the application of the present model to published radiobiological experiment conditions shows that oxygen depletion can only have a negligible impact on radiosensitivity through oxygen enhancement, especially at typical experimental oxygenations where a FLASH effect has been observed.

**Conclusion:** We show that the magnitude and dependence of the "oxygen depletion" hypothesis are not consistent with the observed biological effects of FLASH irradiation. While oxygenation plays an undoubted role in mediating the FLASH effect, we conclude that state-of-the-art radiation chemistry models do not support oxygen depletion and radiation-induced transient hypoxia as the main mechanism.

## Introduction

From the late 1950's on, pioneering studies reported an enhanced radioresistance when using ultra-high dose rates both *in vitro* [*1,2,3,4*] and *in vivo* [*5,6,7*]. In an environment where the radiobiological oxygen effect had been recently discovered [*8,9,10*], the first theories aiming at a mechanistic description of the oxygen-induced radiosensitivity, notably the so-called oxygen fixation hypothesis [*11*], were progressively developing. Parallel and sometimes overlapping investigations [e.g. *12,13*] on both topics, together with the observation of "breaking" survival curves (displaying a slope corresponding to anoxia after a threshold) under low oxygenations at ultra-high dose rate ($10^9$ Gy/s), rapidly put the focus on oxygen depletion as the suspected mechanism behind the increased radioresistance under FLASH conditions [*1*]. In the chemical stage of radiation damage of oxygenated targets, a large part of the hydrated electrons and hydrogen radicals produced by water radiolysis react with the dissolved molecular oxygen through

$$e_{aq}^- + O_2 \rightarrow O_2^{\cdot -} \tag{1a}$$

$$H\cdot + O_2 \rightarrow HO_2\cdot, \tag{1b}$$

producing the cytotoxic superoxide anion and its protonated form, the perhydroxyl radical. Under ultra-high dose rate irradiation, when oxygen is consumed according to reactions (1) and its rediffusion into the irradiated volume can be excluded, a transient acute radiation-induced hypoxia has been hypothesized to increase radioresistance.

A current revival of experimental investigations into the so-called FLASH effect [*14,15,16,17,18,19*] with electron irradiation is successfully demonstrating a (differential) sparing effect *in vitro* and *in* *vivo* when applying ultra-high dose rate (>40Gy/s). Beneficial effects observed include reduced pulmonary fibrosis and pneumonitis [*14*] and cutaneous lesions [*14*,*15*,*17*] and a decrease / absence of late neurocognitive end points [*16*] all of which can potentially widen the therapeutic window between tumor control and normal tissue (NT) complication probability. Discussions about the underlying mechanism [*20,21,22,23,24,25,26,27,28,29*] are, however, ongoing and growing especially in the last two years. Some authors [*21,22*] point towards the early chemical stage of radiation damage as the decisive time frame, guided by the duration of the irradiation itself. Besides radiolytic oxygen depletion (ROD), which is still among the most favored hypotheses [*20,22,23,25,26,29*], it was discussed whether the differential effect could be justified on the basis of a lipid hydroperoxide / peroxyl radical effect [*22,24,30,31,32*], with Labarbe et al. [*28*] offering a very interesting alternative mechanism based on peroxyl radical recombination. In a similar direction, recent molecular dynamics simulations [*33*] indicate that ultra-high dose rate causes formation of clusters of ROS until 1ns which then screen each other from reacting with DNA. Other tentative explanations include a different inflammatory response via the reduced exposure of circulating blood lymphocytes [*20,27*], a decrease in reactive oxygen species production [*16*], differences in redox metabolism in healthy vs. tumor tissue [*22*] and the selective sparing of hypoxic stem cell niches in normal tissue (NT) [*34*].

Here, we tackle the question of radiation-induced oxygen consumption based on first principles. Our novel physico-chemical approach focuses on the chemical evolution of electron tracks (ps-µs after physical interactions) and implies a non-constant oxygen consumption rate. We start from the $pO_2$-dependent track chemistry of 1 MeV electrons in water in assuming a short radiation pulse (~ms range) which effectively prevents a rediffusion of oxygen intra-irradiation from outside the irradiated volume. The simulated $O_2$ consumption yield is used to study the transient variation of target oxygenation. Subsequently, the instantaneous oxygenation under FLASH conditions is combined with the radiobiological oxygen effect through the oxygen enhancement ratio and compared to that in conventional irradiation to identify the dose and oxygenation range where a sparing effect is expected. A tentative application of this general analysis to published pre-clinical [*14,15,16*] and clinical [*17*] FLASH experiment conditions is appended in supplementary material S3. Using an alternative method, we compute *in vitro* cell survival [*18*] under dynamic vs. constant oxygen pressure.

## Materials and Methods

*Monte Carlo simulations*

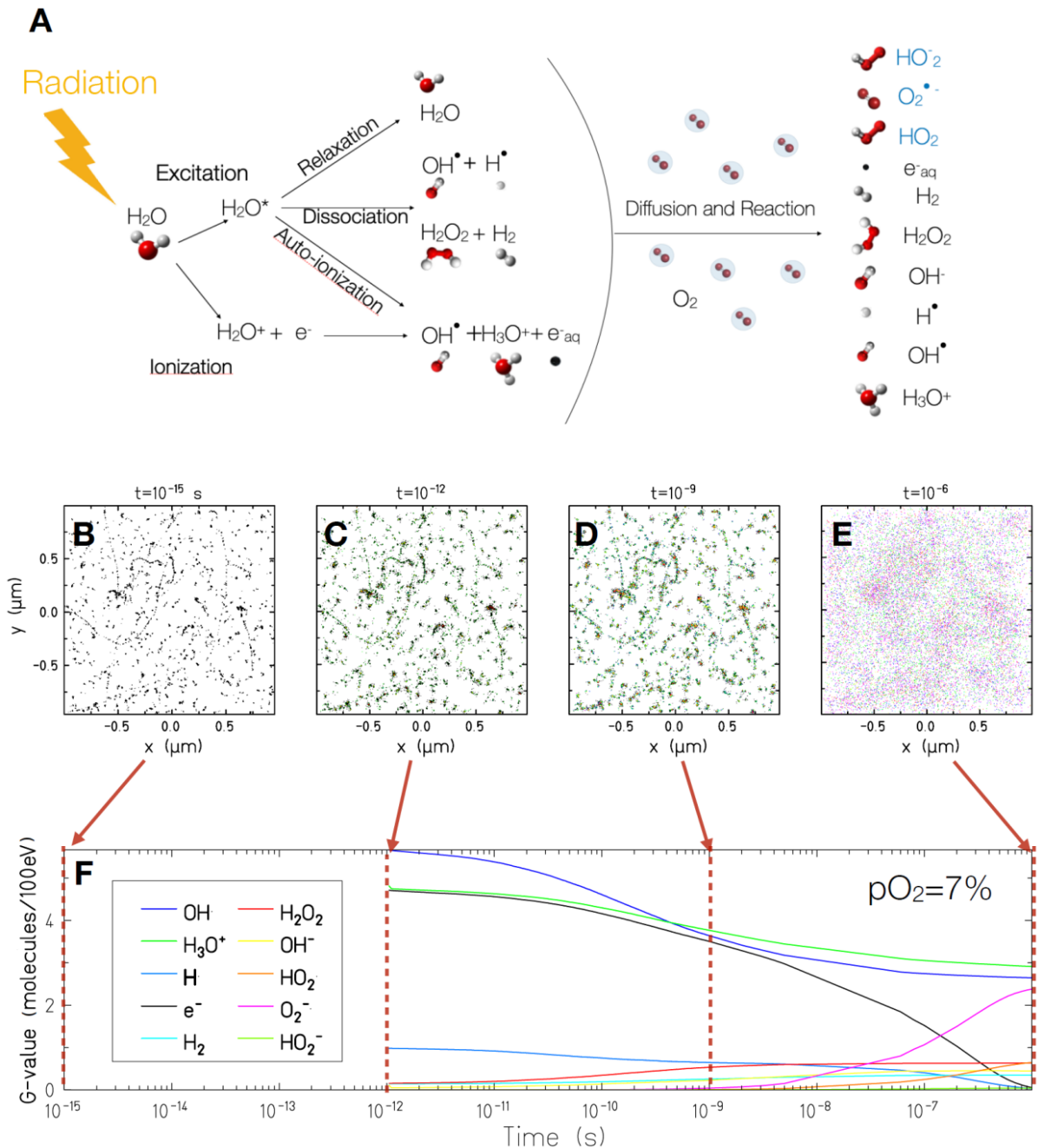

**Figure 1.** Scheme of the physico-chemical electron track model following irradiation of a water target. (A) schematic of physical events and ensuing chemical track evolution from left to right. (B-E) Projection, at different time points, of physical ($10^{-15}$ s) and chemical (1 ps, 1 ns, 1 µs) tracks of 600 1 MeV electrons impinging along the z-axis, corresponding to a dose of 1.49 Gy. (F) Temporal evolution of chemical yields at corresponding times based on single track calculations. Simulation performed with the TRAX-CHEM code in a water target in physioxia (oxygen concentration of 91 µM or 7% $pO_2$).

The evolution of electron tracks in water through the physical, prechemical and chemical stage is simulated using the TRAX [35] code and its extension TRAX-CHEM [36,37]. Implementation of track chemistry in water [36] and the addition of dissolved oxygen [37] have been described in detail elsewhere, and a sketch of the simulation procedure is given in fig. 1. In brief, physical particle interactions are simulated with an event-by-event approach and the resulting positions and ionization or excitation levels of the target molecules ($10^{-15}$ s, fig. 1B) are handed over to the pre-chemical stage where molecular dissociation and thermalization of fragments and sub-excitation electrons takes place. Then (~1ps, fig. 1C) these radiolytic species generated close to the corresponding physical events are tracked as they diffuse and react with each other [*36,37*], water, and oxygen, and transition from an initially heterogeneous, track-structure-determined dynamic (fig. 1D) to a homogeneous solution readily described by species concentrations at ~1µs (fig. 1E). Earlier time points correspond to a denser (less diffused) track pattern but can give correct radical yields from 0.5µs on in anoxic conditions, whereas a later time point facilitates scavenging of $e_{aq}^-$ and H· by low $O_2$ concentrations. To avoid any bias induced by time end point selection of the chemical track evolution, all results are presented with a margin (shaded area of figures) to account for sensitivity when calculating the chemical yields at two alternative time points (0.5 µs, 2 µs, see also S1). Inter-track reactions / recombinations (reactions of the radiolytic species of one electron track with those produced by a neighboring one) are not included in the simulations (see Discussion). The total yield of molecules or radicals of each species per deposited energy is recorded throughout the process as time-dependent G-value curve (fig. 1F) and allows to understand the reaction dynamics in the chemical track. Radiolytic yields for net oxygen consumption ($G_{-O_2}$) and superoxide production ($G_{O_2^{·-}+HO_2·}$) are obtained by simulating many 1 MeV electron passages through a defined water volume ($2\times2\times0.5$ µm$^3$ for physical interactions surrounded by additional water to establish charged particle equilibrium) containing a specific concentration of $O_2$ and normalizing the yields originating in that volume by the energy deposited. The estimated statistical uncertainty of the mean values thus obtained is 2%.

*Dose-dependent species concentrations and OER*

Oxygen depletion and superoxide ($O_2^{·-}+HO_2·$) production in FLASH irradiations can be directly calculated as a function of dose, by numerically integrating the corresponding tabulated $pO_2$-dependent yields (units of mol/J equivalent to M/Gy in water)

$$[O_2] = [O_2]_{ini} - \int G_{-O_2} ([O_2](D)) \, dD \tag{2a}$$

$$[O_2^{·-}+HO_2·] = \int G_{O_2^{·-}+HO_2·} ([O_2](D)) \, dD . \tag{2b}$$

To that end, we assume sufficiently short total irradiation times (<100 ms) in order to exclude rediffusion of oxygen into the target volume; in practice this gives an upper limit for the oxygen depletion (maximizing the FLASH effect) and is equivalent to sealed targets irradiated in conventional mode.

Next, OER-weighted dose under dynamic oxygenation is computed as

$$D_{OER,DYN} = \int OER([O_2](D)) \, dD \tag{3}$$

(in contrast to $D_{\text{OER,CONV}} = D \cdot \text{OER}([O_2]_{\text{ini}})$ ) and represents the most general approach to include oxygen-enhanced radiosensitivity and predict the isoeffective dose under varying oxygenation. We use the definition

$$\text{OER}(pO_2) = D_{\text{anoxia}}/D(pO_2) \big|_{\text{same effect}} \tag{4}$$

and parametrization proposed by Grimes and Partridge

$$\text{OER}(pO_2) = 1 + (\Phi_O/\Phi_D)(1-\exp(-\varphi\, pO_2)) \tag{5}$$

($\Phi_O/\Phi_D = 1.63$, $\varphi = 0.2567$) [*38*] which they fitted to cell killing as the endpoint.

*In vitro cell survival calculation*

In order to assess the impact of the dynamical $O_2$ at ultra-high dose rates on cell survival [*18*], dose response curves for all oxygenations fitted to the linear-quadratic model, $S(D) = \exp(-\alpha D - \beta D^2)$, were taken from the original authors [*18*]. They hold α constant while β accounts for oxygenation and dose rate regime,

$$\beta_{CONV} = \beta_0 \left[1 - \left(\frac{pO_{2,ini}}{20}\right)^{0.4464}\right] + \beta_{20} \left(\frac{pO_{2,ini}}{20}\right)^{0.4464} \tag{6a}$$

$$\beta_{FLASH} = \beta_0 \left[1 - \left(\frac{pO_{2,ini}}{20}\right)^{0.7162}\right] + \beta_{20} \left(\frac{pO_{2,ini}}{20}\right)^{0.7162} \tag{6b}$$

where $\beta_0 = 0.0025$ and $\beta_{20} = 0.0194$ are the quadratic coefficients under anoxia and normoxia, respectively. To separate the effect of oxygen depletion at FLASH dose rates from other influences in the experiment, we follow this fitting strategy but combine the CONV parameters with the dynamic (instantaneous) oxygen concentration $[O_2](D)$ according to eq. (2a) and obtain a dynamic quadratic coefficient

$$\beta_{DYN} = \beta_0 \left[1 - \left(\frac{pO_2(D)}{20}\right)^{0.4464}\right] + \beta_{20} \left(\frac{pO_2(D)}{20}\right)^{0.4464}. \tag{7}$$

The survival under dynamic oxygenation $pO_2(D)$ is determined from

$$S = \int \frac{dS(\beta(pO_2(D)),D)}{dD} dD = -\int \left(\alpha + 2\beta D + D^2 \frac{d\beta}{dD}\right) e^{-\alpha D - \beta D^2} dD \tag{8}$$

and is evaluated iteratively.

## Results

The oxygenation-dependent molecular yield per dose for dissolved oxygen removal for 1 MeV electrons is presented in figure 2. The radiation quality is chosen to be close to the effective energy in a superficial target irradiated with a FLASH-capable accelerator, e.g. a modified clinical linac. In well oxygenated targets, the yield for oxygen removal does not show a dependence on initial oxygenation. In this plateau region the calculated ROD yield of 0.33 µM/Gy (3.23 molecules /100 eV) is in excellent agreement with the literature value of 0.32-0.33 µM/Gy [*22,39,40*] and very similar to that obtained for low-LET protons (0.35 µM/Gy [*37*], see also S4). Below an oxygenation value around 65 µM (5% $pO_2$ at room temperature), the curve exhibits a steep decrease as the sparse distribution of oxygen in the target slows down the chemical track dynamics sensibly and reactions (1) fail to reach equilibrium within one microsecond [37]. Additional reaction pathways can take place in biological media, such as the reaction of the abundant radiolytic OH· radicals to produce organic radicals R·+$H_2O$ which subsequently undergo fast reactions with $O_2$. In accordance with experimental ROD values in cell culture medium [41,42], a tentative oxygen removal yield curve scaled by a factor 4/3 is therefore included. Note that this applies strictly only to the plateau region, since the different kinetics may affect the results. In what follows, the radiolytic yields in water at 1 µs will be regarded as the reference for further calculations, while the shaded area is propagated to investigate the impact of the time end point and liquid chosen.

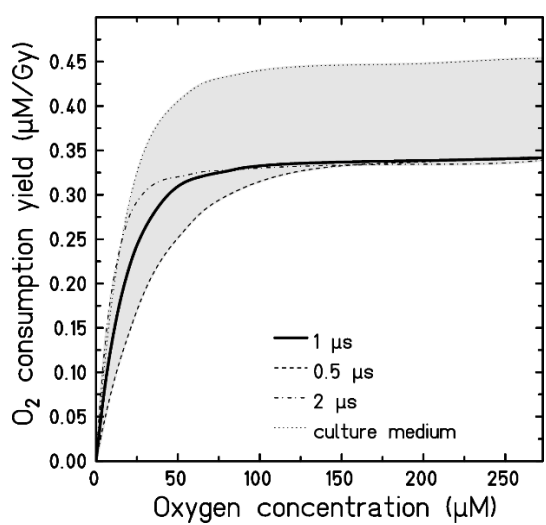

**Figure 2.** Calculated oxygenation-dependent radiolytic yields for ROD at different time points around the end of the track (heterogeneous) chemistry stage (1 MeV electron radiation). The dotted curve approximating cell culture medium is obtained by scaling the 1µs TRAX-CHEM simulation results by a factor 4/3.

In fig. 3A, the oxygen concentration is shown as a function of dose for different initial oxygenations $pO_{2,ini}$ covering a typical range from physioxic NT (7%, 5%) to hypoxic tumors (1%) [43]. In the dose range depicted, oxygen consumption is proportional to dose, and complete "depletion" cannot be observed for any $pO_{2,ini}$.

However, as proposed recently [23], radiation-induced hypoxia reduces radiosensitivity [8,9,46] as described by OER [38,45]. Assuming that the radiobiological oxygen effect is a result of the oxygenation level at the moment of irradiation [12,13,46], the dose weighted by the instantaneous OER($pO_2$), $D_{OER,DYN}$, is then calculated. Results in fig. 3B reveal that similar to ROD, any noticeable effect (deviation from initial slope $D·OER(pO_{2,ini})$, representative of constant OER) arises at high doses (generally well over 100 Gy, see S2). Moreover, the present calculations yield a smaller ratio $D_{OER,DYN}/D_{OER,CONV}$ (at a given absorbed dose) for the low oxygenations typically found in hypoxic tumors, but unrealistic for the surrounding healthy tissue.

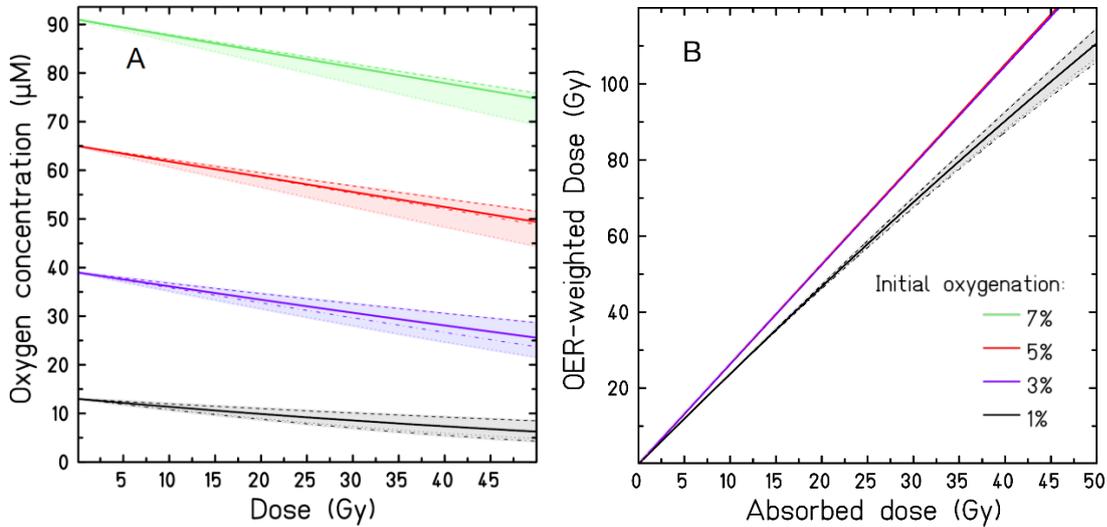

**Figure 3.** Dose dependence of oxygen concentration under FLASH irradiation conditions with 1 MeV electrons for different initial oxygenations (A) and its implication on, OER-weighted dose $D_{OER,DYN}$ (B) for a short pulse, at room temperature. Note that in (B), the lines for 3%, 5% and 7% initial $pO_2$ overlap. While concentration is given in absolute SI units, the initial target oxygenations corresponding to each line are given in % $pO_2$ for ease of comparison to experiments.

To investigate this closer, fig. 4 depicts the relative $D_{OER}$ of FLASH with respect to conventional irradiation as a measure for the FLASH sparing effect. Target oxygenation is here used as a proxy for normal/tumor tissue nature. For typical NT oxygenations (5–7% equiv. to 65–91 µM), the $D_{OER}$ ratio remains close to one (no sparing) for elevated hypothetical doses up to ~150 Gy and displays a slow decrease afterwards. Intermediate oxygenations (2-3% equiv. to 26–39 µM) show a rather flat behavior of $D_{OER,DYN}/D_{OER,CONV}$ for low doses and a ratio of 0.98–0.93 at 100 Gy. Initial oxygenations of 1% (13 µM) and lower lead to steeper, nearly linear slopes starting from 0 Gy. The maximum sparing due to radiation-induced transient hypoxia for a 25 Gy dose is $D_{OER,DYN}/D_{OER,CONV} = 0.94$, achieved with oxygenations of 0.3–0.5% (3.9–6.5 µM); for even lower oxygen concentration, the corresponding curves become shallower again (less sparing). This is due to a tradeoff effect between oxygen consumption and oxygen availability for further depletion.

The relative effect plotted in fig. 4 easily identifies the oxygenations leading to a better FLASH sparing for a given dose. Specifically, in order to afford a better sparing in the more oxygenated volume, it is necessary that the corresponding $D_{OER,DYN}/D_{OER,CONV}$ curves cross each other. In the dose range presented, this occurs only for a few combinations as follows. 2% and 0.1% and 1% and 0.3% curves intersect around 150 Gy and the 1% and 0.1% curves at ~40 Gy. None of these oxygenation levels are considered physioxic, i. e. a typical oxygenation level found in NT [43].

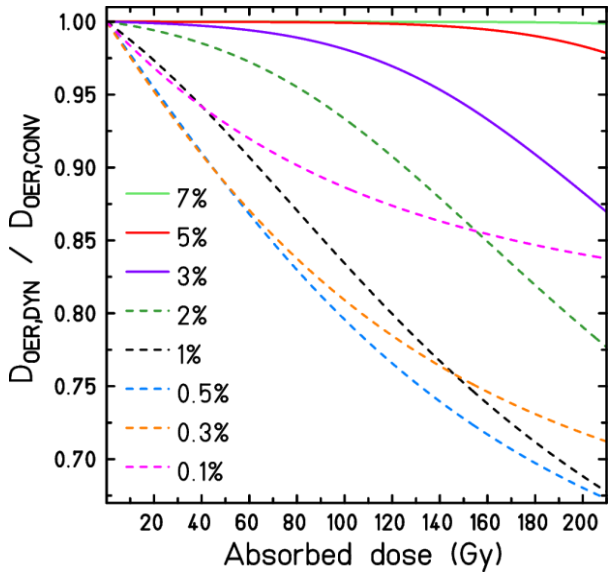

**Figure 4.** Relative OER-weighted dose (ratio) of FLASH vs. conventional dose rate irradiation for different initial oxygenations as a function of dose (1 μs time end point in water). Solid lines indicate the typical range of normal tissue oxygenations, while dashed lines represent different degrees of hypoxia found in tumors.

Finally, the conditions of the recent *in vitro* experiment at ultra-high dose rates of Adrian et al. [18] are reproduced. They reported prostate cancer cell survival for different starting oxygenations in FLASH vs. conventional electron beam irradiation, see fig. 5, which we complement with dynamic $O_2$ concentration for the different $pO_{2,ini}$. For all of them, the resulting survival $S_{DYN}$ (fig. 5) is slightly higher than the experimental CONV curve but falls short of the experimental FLASH results (solid curves) which indicate a much more pronounced sparing effect. The numerical ratios $S_{DYN}/S_{CONV}$ for 18 Gy as compared to the experimental ratio $S_{FLASH}/S_{CONV}$ [18] are listed in table 1. For cell survival of prostate cancer cells as an endpoint and a dose of 18 Gy, oxygen depletion is responsible for an increase in survival of around 10% (a maximum of 22% in the most favorable combination) but cannot justify the full 30-490% difference observed experimentally [18].

**Table 1.** Calculated ratio $S(D)_{DYN}/S(D)_{CONV}$ obtained at 18 Gy. The stated values are derived using 1 μs yields in water, with the range of possible ratios for the other combinations from fig. 2 given in brackets.

| $pO_{2,ini}$ | $S_{DYN}/S_{CONV}$ | $S_{FLASH}/S_{CONV}$ exp. [18] |
|---|---|---|
| 8.8 % | 1.06 (1.06-1.12) | 1.3 |
| 4.4 % | 1.09 (1.07-1.12) | 4.9 |
| 2.7 % | 1.10 (1.08-1.14) | 3.27 |
| 1.6 % | 1.11 (1.07-1.22) | 3.15 |

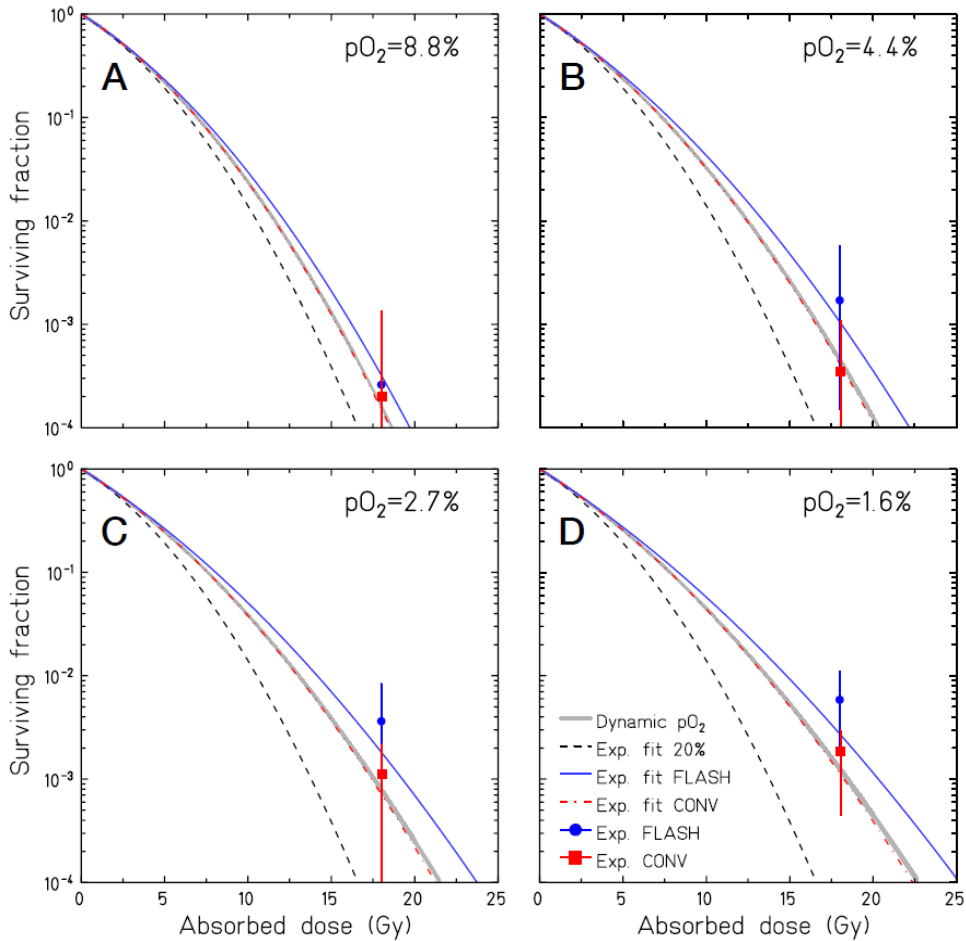

**Figure 5.** Impact of oxygen consumption under FLASH conditions on the survival curves of prostate cancer cells. Lines "Exp. fit CONV" and "Exp. fit FLASH" show the surviving fraction $S(D)$ of prostate cancer cells at different oxygenation levels for conventional (red) and FLASH (blue) irradiations fitted by Adrian et al. [18] to their *in vitro* data. The experimental data points at 18 Gy are reproduced from [18]. For comparison, the shaded areas ("dynamic $pO_2$") represent the model prediction of cell survival $S_{DYN}$, accounting for the oxygen consumption results. Note that in (A), the conventional fit curve and calculated survival $S_{DYN}$ overlap partially. As in the figs. 2 and 3 of the main article, the shaded areas contain the range of results for the different radiolytic yield time points and the underlying target medium.

## Discussion

With this novel, radiation chemistry-based modelling approach, we find that full oxygen depletion has to be discarded as the sole explanation for increased NT sparing under FLASH irradiation conditions due to the large doses required. Furthermore, the oxygenation-dependent difference in $D_{OER,DYN}/D_{OER,CONV}$ translates into a narrowing of the therapeutic window due to selective sparing (if any) at typical tumor oxygenations. In Fig. 4, no realistic combination can be identified where the FLASH sparing due to radiation-induced transient hypoxia is expected to favor physioxic target volumes. The calculation corresponding to FLASH *in vitro* [18] predicts only slightly higher clonogenic survival under FLASH conditions and therefore does not support oxygen depletion as the (principal) cause for the much larger difference observed experimentally.

We focus on a situation equivalent to exposing a sealed target volume at conventional dose rate (to impede rediffusion); a reasonable approximation for ultra-high dose rates which serves as an upper limit condition for ROD. The chemical difference experimentally observed between these two conditions [47] gives valuable hints towards the true mechanism at work, particularly a concentration-dependent homogeneous radical effect (taking place after track dissolution and possibly linked to oxygen, e. g. as described in [28]) which affects the biological response. Chemical inter-track effects (in the heterogeneous stage), in contrast, are not likely to take place. Most prominently, they affect OH· radicals and their recombinations [48,49,50] but also reduce oxygen consumption due to the competition of reactions (1) with recombination among the primary radicals [51,52]. The impact starts to be noticeable e. g. for 10 Gy delivered instantaneously [48]. For typical parameters used in recent FLASH experiments this is not the case, see also fig. 1C-1D which is representative of a single pulse given *instantaneously*. The situation is, however, very different for historical results showing an enhanced radioresistance *in vitro* [1,2,3,4] where dose rates as high as $10^9$ Gy/s were used and irradiations were accomplished in 3 ns.

While the present ab initio approach links heterogeneous stage effects to the relevant macroscopic, experimental results [39,40], some limitations should be kept in mind and reduce the applicability of these simulations in water to living cells or tissues. First, water is a convenient target medium for simulations, but a biological target is different in many aspects, such as the existence of cellular targets sensitive to chemical damage, reactions of radiolytic species with target molecules competing or adding up with those in water, the buffered pH condition (different in tumor / NT), the presence of catalytic enzymes to remove cytotoxic species, etc. As pointed out by Wardman [53], even primary radical interactions (notably of $e_{aq}^-$ and H·) are likely to change in presence of high concentrations of competing biomolecular scavengers [54]. This would cause the dissolved oxygen to react in the next stage with the organic radicals R· formed and produce ROO· partially in peroxidation chain reactions [30,28]. A meaningful quantification of the latter organic peroxyl radical yield (and associated oxygen consumption), and of possible subsequent radical-radical reactions, needs however to be performed with a simulation based on homogeneous species concentrations. This goes beyond the focus of the present study but could be integrated into our code in future developments. In a biochemically more realistic model, it would also be desirable to extend the radiation chemical simulations until longer time scales, where the oxygen effect becomes effective biologically (a few ms [12]). In water, a longer end time would entail differences mainly for the tumor oxygenation range. In tissue, even if the responsible reactions may change from the water case, the overall ROD does not seem to be higher [*47*]. In order to serve as a reference for future modeling work, a larger body of experimental radiation chemical results in complex target media is highly needed.

Finally, OER here used to translate an oxygenation change to a modification in effective dose, is a concept adopted from extensive *in vitro* evidence focusing mostly on clonogenic cell survival, but is potentially not suited for predicting the late effects in healthy tissue seen *in vivo*. Global and possibly indirect effects manifesting at longer time scales post-irradiation may need to complement the mechanistic explanations of the classical oxygen effect which center on a chemical enhancement of localized DNA lesions involving damage fixation or translocation within a defined ms time window directly by reaction with oxygen.

Our findings agree with some inconsistencies of the oxygen depletion hypothesis noted earlier [*14,20,28*] and warnings about possible larger effects in hypoxic tumors than in physioxic tissues [*23,24,18*]. In particular, Pratx and Kapp [*23*] applied a radiolytic oxygen depletion model with constant depletion rate, but accounting explicitly for diffusion, blood vessel perfusion and metabolic oxygen consumption, to conclude that oxygen depletion yields an OER modification only for hypoxic starting conditions. Full depletion of oxygen is found unlikely for a typical 30 Gy dose and very low oxygenation (0.01%) with a GPU-based Monte Carlo calculation [*52*]. A recent article by Petersson et al. [*26*] presents a simplified mathematical framework to describe oxygen variation including reoxygenation during irradiation which can be adjusted to reproduce some experimental findings under FLASH conditions [*16,18*]. The authors acknowledge, however, that many of the key response parameters remain uncertain; this circumstance compromises that model's capability to predict the magnitude of the effect in experimental conditions. Finally, oxygen consumption would show no large dependence on intra-pulse dose rate if the total irradiation time is sufficiently short, however a delicate dependence on pulse dose, length and frequency has been established *in vitro* and pre-clinically [*56,55*].

## Conclusions

Based on the chemical track evolution of electrons in oxygenated water, and allowing for a margin of conditions (time end point for chemical yields and target medium) around the experimentally confirmed [*39,40*] 0.33 µM/Gy depletion in buffered water, we have shown that oxygen removal in water does not account for a sparing effect at FLASH dose rates and clinically realistic doses. A full depletion is not supported by the present computational model, and no substantial sparing effect emerges when applying calculated cell survival and OER-weighted dose to typical experiment conditions. Indeed, based on the model predictions a differential effect would arise and provide better protection for poorly oxygenated targets than for those ranging from fully oxygenated to physioxic.

The present results, thus, lend support to the growing awareness that further and more complex mechanisms should be investigated beyond the simple ROD hypothesis.

# Supplementary material

## S1. Methods details

*Monte Carlo simulation time points*

It is generally observed and accepted that the chemical track structure of a single incident particle of radiation reaches the spatial dissolution into a homogeneous distribution at ~1µs. For a pure (anoxic) water target, this time end point is also where chemical equilibrium is attained, radiolytic yields (G-values) are stable, and TRAX-CHEM results have been benchmarked against the available experimental data. In presence of oxygen (a target molecule in homogeneous dissolution, according to the given concentration), further reactions with the radiolytic species can still take place and produce changes in species yields even later than 1 µs which may affect the present results. The authors therefore include within the present results two more time points which we consider delimit the meaningful range for biological applications: on the one hand, 0.5 µs is the earliest moment where the radiolytic yields generally adopt their final value *in anoxic conditions*, thus representing a stable situation after radiolysis where track chemistry has lost its prevalence. On the other hand, the selection of 2µs is an extended time point guaranteeing an even better species homogeneity, especially when adding $O_2$, but limits the possible reaction partners to $H_2O$ and $O_2$ at a stage when other competing reactions with additional solutes of a realistic biological environment (e.g. a cell) become very important. For simulation over longer times, an inclusion of specific target molecules, especially those likely to suffer chemical damage from the radiolytic radicals as well as radical scavengers / catalyzers capable of counteracting, would be mandatory and give additional interesting insights into the biochemical radiation effects. This functionality is planned but not currently included in TRAX-CHEM or any other tool available to the authors. In all results figures, the shaded areas depict the range of variation of the presented results when using different time points (0.5µs to 2µs) of the underlying radiolytic yields or changing the target medium.

*Acid-base equilibrium*

In the present track structure approach, $O_2^{\cdot-}$ is almost exclusively produced by aqueous electrons and $HO_2\cdot$ by $H\cdot$, so that the predicted ratio of both therefore closely follows the yield of $e_{aq}^-$ vs. $H\cdot$. In a biological environment, the relative abundances of this acid-base pair would rapidly equilibrate depending on pH. In order to avoid confusion, we limit the presentation of results and discussion to the sum of both.

*Concentration units*

With the objective to use absolute SI units for oxygenation which are independent of solubility, we opt for µM for oxygen concentration $[O_2]$ in graphs and numerical results. However, for a better comparability with experimental data, the oxygenation levels of particular target examples is sometimes also described in partial oxygen pressure (% $pO_2$), and the definition of oxygen enhancement ratio is based on $pO_2$ following usual practice. The corresponding relative oxygen saturation in experimental targets was generally calculated according to Henry's law for a temperature of 25°C where 21% $pO_2$ = 273 µM. Only for the reconstruction of *in vivo* conditions, Henry's constant was set to its value at 37°C, $H^{cp}=1.0264\cdot 10^{-5}$ mol/(m³Pa), so that 21% $pO_2$ corresponds to 218 µM. The chemical diffusion-reaction model itself remained unchanged from Boscolo et al. [37].

*OER convention*

Note that the OER definition followed here is different from the ratio $D(pO_2)/D_{normox}$ which has been adopted in some studies and rather reflects an "oxygen diminution ratio" (ODR); OER=1/ODR · ($D_{anoxia}/D_{normox}$).

## S2. Species concentrations and their dose dependence

*ROS production: superoxide and its protonated form*

Since in oxygenated targets a larger impact of oxidative stress is expected, the production of superoxide $O_2^{\cdot-}$ and its conjugate acid perhydroxyl $HO_2\cdot$, which are closely related to radiolytic oxygen depletion in water, are obtained as an example of changes in ROS production. Oxygenation-dependent chemical yields per deposited dose for $O_2^{\cdot-}/HO_2\cdot$ production by 1 MeV electrons are presented in figure S1. In well oxygenated targets, the yields for oxygen consumption and production of $O_2^{\cdot-}/HO_2$ do not depend on initial oxygenation. Below an oxygenation value of around 65 µM (5% $pO_2$ at room temperature), the curve however exhibits a steep decrease as the sparse distribution of oxygen in the target slows down the chemical track dynamics sensibly and reactions from eq. (1, main article) become less frequent.

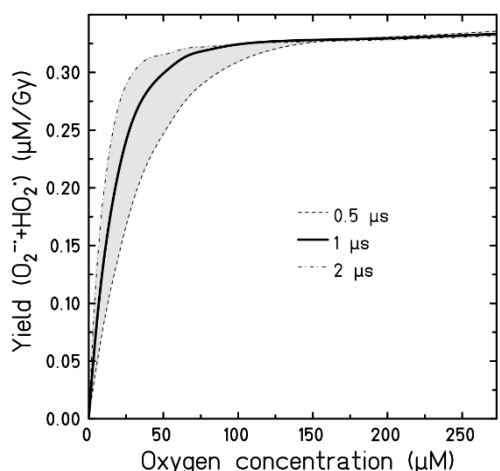

**Figure S1.** Calculated oxygenation-dependent radiolytic yield for the production of superoxide + perhydroxyl at different time points around the end of the track (heterogeneous) chemical stage (1 MeV electron radiation).

*Radiolytic oxygen depletion for high doses*

Even if irradiation with doses exceeding some tens of Gy is usually too slow to comply with the ms dose delivery we assume as a condition, and also clinically irrelevant, we want to present an extended dose dependence analysis until reaching full depletion. Please note that the results of this section represent an approximation and do not account for oxygen rediffusion in the irradiated target, and thus show an upper limit to oxygen consumption. In an extended dose range (fig. S2A), oxygen consumption for different initial oxygenations $pO_{2,ini}$ is found to be not proportional to dose any more (as often approximated). The successive deviation from linearity is especially pronounced for the low range of oxygenation typically found in tumors [43]. Absorbed doses required to deplete virtually all dissolved oxygen range from ~400 Gy ($pO_{2,ini}$ = 7%) to ~150 Gy ($pO_{2,ini}$ = 1%), far exceeding preclinically accepted doses in a single fraction even at ultra-high dose rate [44]. The dynamical oxygen depletion introduced here is fundamental to estimate the high doses which would be necessary. Simultaneously to progressive oxygen consumption, superoxide production is steady until a complete oxygen depletion is approached (fig. S2C). The resulting plateau is observed at 89.6 µM (7 % $pO_{2,ini}$), 64.0 µM (5 % $pO_{2,ini}$), 38.4 µM (3 % $pO_{2,ini}$), and 12.8 µM (1 % $pO_{2,ini}$), respectively.

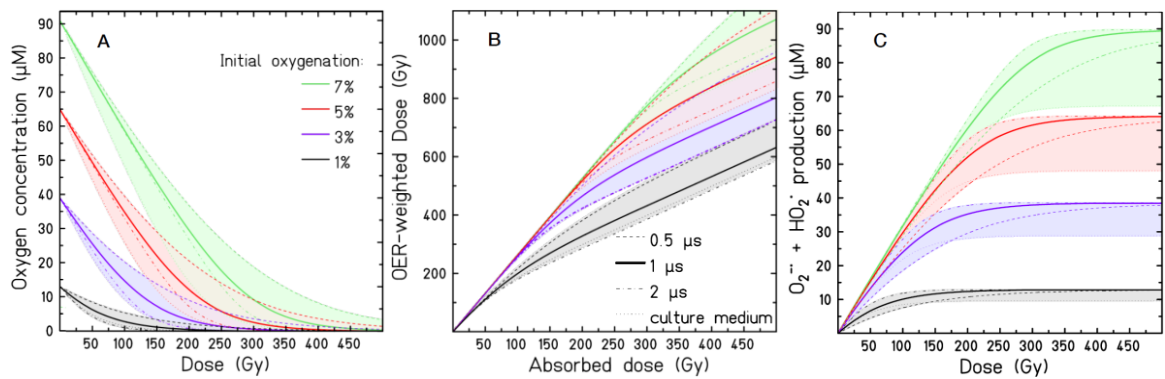

**Figure S2.** Chemical species concentrations under FLASH irradiation conditions with 1 MeV electrons for different initial oxygenations and its implication on the radiobiological outcome for high doses. (A) Radiochemical oxygen consumption with dose, (B) production of superoxide + perhydroxyl, and (C) OER-weighted dose $D_{OER,DYN}$ for a hypothetical short pulse, all at room temperature. The initial target oxygenations corresponding to each line are given in % $pO_2$ for an easier comparison to experimental data.

*Relative dose-dependent superoxide production, FLASH vs. conventional irradiation*

The production of cytotoxic superoxide is the second end point influenced directly by oxygen pressure in the heterogeneous (track) chemistry stage of radiation damage. In addition to radiolytic oxygen consumption, the reduced superoxide production $O_2^{·-}+HO_2·$ of FLASH with respect to conventional irradiation conditions can therefore serve as a proxy for FLASH tissue sparing. The relative effect is presented in figure S3. It shows, indeed, a generally larger sparing effect at the same dose than the effective dose $D_{OER}$ discussed in the main article, indicating higher sensitivity to transient hypoxia at ultra-high dose rate. However, the intrinsic nature of superoxide production dictates that the less oxygen is available, the less superoxide will be produced, and the relative curves never cross to predict a selective NT sparing – hypoxic tissues are still favored over physioxic ones.

In water, the ROS produced can undergo spontaneous disproportionation giving hydrogen peroxide at longer time scales in the chemical evolution, so that a decreased production might be linked to a decrease in $H_2O_2$. For a quantitative analysis of that aspect, other competing reactions (e.g. with OH·) need to be included in simulations of time points extending into the homogeneous chemistry stage.

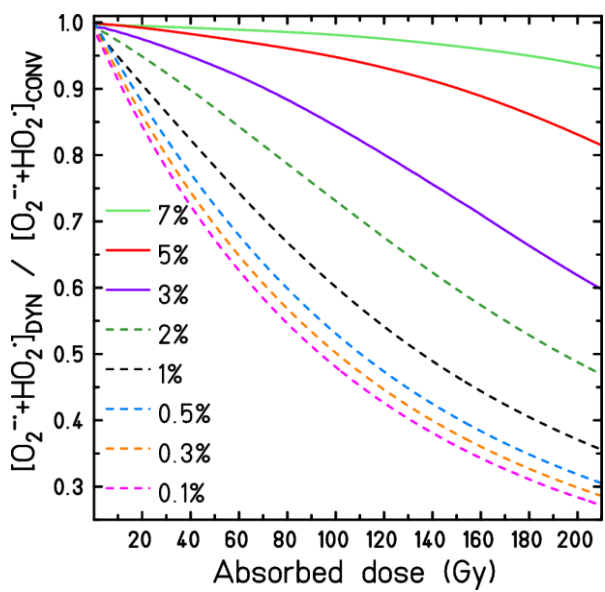

**Figure S3.** Relative superoxide ($O_2^{·-}$+$HO_2·$) production of FLASH vs. conventional dose rate irradiation for different initial oxygenations as a function of dose (calculated for 1 µs time end point in water). Solid lines correspond to typical normal tissue oxygenations, while dashed lines represent different degrees of hypoxia found in tumors.

## S3. Application to pre-clinical/clinical experiment conditions

Finally, to enable a quantitative comparison, we present numerical values for the specific experimental conditions of recent in vivo studies [14,15,16,17] where significant sparing was reported in FLASH mode. This represents a tentative application of the present model to experiments realized in biological organisms and needs to be interpreted with precaution, c. f. Discussion in main article. Summarizing these studies, Favaudon et al. [14] observed a reduced induction of pulmonary fibrosis, pulmonitis and cutaneous lesions in a syngeneic, orthotopic murine lung tumor model by FLASH irradiation as compared to conventional dose rate. The lung sparing was quantified by a dose modifying factor of 2 for the same average late effect score in the recent review of Vozenin et al. [20], while antitumor efficiency and survival was comparable in both modalities. In a dose escalation study comprising doses between 22-41 Gy, Vozenin et al. [15] showed significant healthy skin sparing (≥ 20% extra dose in FLASH for fibronecrosis end point) in cats and a minipig, in combination with good tumor control, afforded by the possibility of giving higher doses, in the cat patients. Montay-Gruel et al. [15] found a marked decrease (up to absence for some end points) in late neurocognitive effects after a 10 Gy dose when using FLASH vs. conventional dose rate in mice. Furthermore, a first human skin lymphoma was treated with FLASH radiotherapy [17] safely for the first time and yielded a very promising estimated effective FLASH dose to healthy skin of 2/3 of the absorbed dose.

Table S2 presents our model results reproducing the experimental conditions of several parts of the above mentioned experiments. This includes the murine lung tumor and healthy lung from [14] at the 17 Gy dose (data available for both conventional and FLASH with clearly diverging late effect scores [21]), minipig and cat skin sparing ≥ 20% from Vozenin et al. [15] with good tumor response as a secondary endpoint, mouse whole brain irradiations to 10 Gy which showed neurocognitive benefits at FLASH dose rates [15], and the first human patient case [17] treated with 15 Gy (cutaneous lymphoma and healthy skin). The tumor xenograft results (tumor control) also presented in [14] are not modeled since the oxygenation would be very difficult to estimate, highly inhomogeneous, and very imprecise. Since tissue-specific oxygenation is not stated in the original studies, we rely on a review by McKeown et al [43] on average oxygenation values for a range of tumors and the corresponding NT. Target oxygenations were set by selecting the closest available match for each case, from 3.4% to 5.9% in normal tissue and from 1.5% to 2.1% in tumor. Calculated production of the superoxide anion and its protonated form, and OER-weighted dose are given for conventional irradiation together with the corresponding % values (FLASH/ conventional) and final oxygen pressure $pO_{2,fin}$ for FLASH mode. Results for normal tissue and tumor are listed beneath each other. It is evident from these normalized values that no significant sparing effect is predicted for healthy tissue. $D_{OER,DYN}$ = 100% for all normal tissues but presents slight drops (maximum of 2.3%) for tumors, smaller and/or opposite to the desired differential effect [14,15,16,17]. Interestingly, superoxide + perhydroxyl production is more sensitive to the transient change in oxygenation and displays a maximum difference FLASH vs. CONV of 3.4% in normal tissue and up to 11.4% in tumor. Even if these magnitudes are more noticeable, they predict, again, a differential sparing of the tumors in disagreement with [14,15]. The present results therefore fail to explain the experimental findings [14,15,16,17] through oxygen depletion or a decrease in ROS production. Quantitative predictions based on oxygen depletion analysis and OER are therefore in sharp contrast to most of the experimental findings at FLASH dose rates so far [14,15,16,17,18] where a differential effect in favor of NT sparing has been shown.

**Table S1.** Model results for superoxide production, final oxygenation and OER-weighted dose in conditions employed in different pre-clinical and one clinical experiment. For the FLASH results (right block of columns), the range given in brackets indicates the confidence interval of values when applying the different sets of chemical yields from figs. 2 and S1 (shaded areas).

| Experiment | Dose (Gy)[a] | $pO_{2,ini}$ (%)[b] | Conventional dose rate | | FLASH | | |
|---|---|---|---|---|---|---|---|
| | | | $[O_2^{\cdot -}+HO_2\cdot]$ (μM) | $D_{OER,CONV}$ (Gy) | $[O_2^{\cdot -}+HO_2\cdot]$ (% of conv.) | $pO_{2,fin}$ (%) | $D_{OER,DYN}$ (% of conv.) |
| Mouse whole brain [16] | 10 | 3.4 | 2.78 | 26.3 | 98.2 (97.4-99.3) | 3.13 (3.10-3.20) | 100 |
| Minipig skin [15] | 31 | 5.3 | 9.48 | 81.5 | 98.1 (96.6-99.4) | 4.39 (4.09-4.54) | 100 |
| Cat, healthy skin/ mucosa [15] | 33 | 5.9 | 10.3 | 86.8 | 98.1 (96.5-99.4) | 4.91 (4.59-5.05) | 100 |
| Cat squameous cell carcinoma [15] | 33 | 1.9 | 7.22 | 85.5 | 88.6 (88.6-90.0) | 1.27 (1.10-1.48) | 98.5 (97.7-99.1) |
| Mouse lung [14] | 17 | 5.6 | 5.24 | 44.7 | 99.2 (98.5-99.7) | 5.09 (4.92-5.17) | 100 |
| Lung tumor [14] | 17 | 2.1 | 3.84 | 44.3 | 96.1 (96.1-97.1) | 1.74 (1.63-1.85) | 99.3 (99.3-99.5) |
| Human patient, healthy skin [17] | 15 | 5.3 | 4.59 | 39.4 | 98.9 (98.2-99.7) | 4.86 (4.71-4.93) | 100 |
| Human skin lymphoma [17] | 15 | 1.5 | 2.77 | 38.2 | 96.8 (96.1-98.0) | 1.24 (1.16-1.33) | 99.1 (98.6-99.3) |

[a]For dose escalation studies, a representative intermediate dose was chosen.
[b]Initial oxygenation is estimated based on the values given for certain tumors and normal tissues in [43]

## S4. Use of protons or ions for FLASH radiotherapy

Regarding the possibility to use protons or ions for FLASH radiotherapy [53], considering the present results together with [O$_2$]-dependent yields for ions at different LET from [35,54] allows to conclude that high energy protons (corresponding to the beam in the entrance channel) present oxygen consumption characteristics numerically very close to the electrons discussed here. Low energy protons with higher LET (close to stopping in the target volume), however, deplete less oxygen per dose so that the differences found in superoxide production and OER would be shifted to higher doses than for electrons, or low LET radiation in general. Likewise, heavier ions such as C exhibit even smaller yields $G_{-O_2}$, $G_{O_2^{\cdot -}}$, and $G_{HO_2\cdot}$ due to prominent early intra-track species recombination. Consequently, compared to the electrons investigated here, considerably higher doses would be needed for complete oxygen depletion or a significant impact on superoxide production.

## Acknowledgments

The authors thank K. Petersson for providing numerical cell survival data from [18].